\documentclass[showpacs]{revtex4}
\usepackage{amssymb}
\usepackage{graphicx}
\usepackage{dcolumn}
\usepackage[intlimits]{amsmath}
\usepackage{hyperref}

\begin{document}

\title{Exact Energy Levels and Eigenfunctions of an Electron on a Nanosphere
Under the Radial Magnetic Field}
\author{A. \c{C}etin}
\affiliation{Department of Physics, Faculty of Sciences and Arts, Kilis 7 Aral\i k
University, 79000 Kilis, Turkey\\
E-mail address: abdurahmancetin@gmail.com\\
}

\begin{abstract}
The exact energy levels and wave functions of an electron free to move on a
sphere under the radial magnetic field is found. Wave functions are
expressed in terms of Jacobi polynomials which were well-defined and have
orthogonality relation, recurrence relations, series expansions etc. We also
discussed the the wave functions and energy levels in case of very large
magnetic field. Landau energy levels are shown for strong constant magnetic
fields occurring on two-dimensional flat surfaces, if the radius is very
large. The results compared with those previously found in the literature.
\end{abstract}

\maketitle
\date{\today}

\section{Introduction}

Flat, two-dimensional systems are carried out on helium and quantum wells.
Similarly, curved two-dimensional systems can be created by both on helium
and solid state systems. In the first case, the multi-electron bubble in the
liquid helium and the second example is the metallic nanoshell \cite%
{Tempereetal2007}.

The electronic properties of curvilinear surfaces have attracted much
interest since the discovery of carbon nanostructures. The motion of charged
particles on a sphere under constant magnetic field have been studied by
authors for discuss the fractional quantum Hall Effect \cite{Haldane1983,
Fanoetal1986, Alaverdianetal1997, YiangAndSu1997}, weak magnetic field and
strong magnetic field properties \cite{Kimetal1992, Aristov1999(1)}, Landau
Levels \cite{AokiAndSuezawa1992, Aristov1999(2)} and optical properties \cite%
{GokerAndNordlander2004}.

Goddart and Olive \cite{GoddardAndOlive1978} gave a detailed description of
the methods of creating radial magnetic fields (Dirac magnetic monopol and
Dirac string). The energy levels of the charged particle on a sphere under a
radial magnetic field was investigated by Ralko and Truong in classical and
quantum mechanic regimes. They obtained solutions in Heun Functions, which
is a generalization of the Gauss Hypergeometric Function. In one of the
coefficients of the Heun equation, they found a condition that led to the
quantification of energy levels \cite{RalkoAndTruong2002}.

While Haldane \cite{Haldane1983}, Ralko and Truong \cite{RalkoAndTruong2002}
determine the single particle wave functions, they take the magnetic vector
potential as $\overrightarrow{A}=-\frac{\hslash S}{eR}\cot \theta ~%
\widehat{\phi }$ due to the radial magnetic field. In this study,
the magnetic vector potential determined in an efficient manner in
section 2. It has been assumed that there is no radial dependence
of the movement of the electron on the sphere and it depends on
only two angle variables. By exactly solving the time independent
Schr\"{o}dinger equation, we found the exact energy eigenvalues
and the wave functions of an electron on a two-dimensional
spherical surface under a radial magnetic field. The
eigenfunctions have been expressed in terms of Jacobi polynomials
that has orthogonality and recurrence relations. In the absence of
magnetic field, it has been shown
that eigenfunctions are reduced to Legendre polynomials. In the $%
p\rightarrow \infty $ limit, it is shown eigenfunctions are reduced to
Laguerre polynomials. Furthermore, the limit $R\rightarrow \infty $ is
discussed, in which case the sphere surface can be taken as a flat surface,
and energy levels have been shown to turn into two-dimensional Landau energy
levels. In the last part, the results have been discussed.

\section{Nanosphere Under the Radial Magnetic Field}

The electrons on a sphere are strongly bound perpendicular to the surface,
while they move freely in directions parallel to the surface. On a sphere,
this means that the full (three-dimensional) wave function describing such
electrons should be factorizable in a function depending only on the angles
and a function depending only on the radial distance. The system can be
considered two-dimensional if all the electrons have the same radial
dependence of their wave function, and if the energy required to change the
radial mode is much larger than the other relevant energy scales involved.

Electrons that can move freely on the sphere are in single-particle angular
momentum states. For a rigid sphere of radius $R$, the energy of a single
electron confined to the surface is called the rigid rotator energy $E_{\ell
}=\frac{\hslash ^{2}}{2m^{\ast }R^{2}}\ell (\ell +1)$, where $\ell $ is the
angular momentum quantum number, $m^{\ast }$ the mass of the electron, and $%
\hslash $ is Planck's constant.

The Schr\"{o}dinger equation of an electron on a metallic nanosphere under
the radial magnetic field

\begin{equation}
\frac{1}{2m^{\ast }}\left[ -i\hslash \boldsymbol{\nabla }+\frac{e}{c}%
\boldsymbol{A}\right] ^{2}\psi =E\psi  \label{1}
\end{equation}

Where $\boldsymbol{A}$ is the vector potential on the sphere due to the
radial magnetic field, $m^{\ast }$ and $e$ effective mass and charge of the
electron respectively and $c$ is speed of light.

Using spherical polar coordinates ($r$, $\theta $, $\phi $) we expect by
symmetry to be able to find a vector potential $\boldsymbol{A}(\theta
)=A(\theta )~\widehat{\phi }$, $\widehat{\phi }$ being a unit vector in the $%
\phi $ direction. The magnetic flux through a circle, C, corresponding to
fixed values of $R$ and $\theta $, and $\phi $ ranging over the values $0$
to $2\pi $, is given by the solid angle subtended by C at the origin
multiplied by $\frac{1}{4\pi }\int \boldsymbol{B}.d\boldsymbol{S}$, namely $%
\frac{(1-\cos \theta )}{2}\int \boldsymbol{B}.d\boldsymbol{S}$. The total
magnetic flux from the sphere surface is $\Phi =\int \boldsymbol{B}.d%
\boldsymbol{S}$ \cite{GoddardAndOlive1978}. Consequently:

\begin{equation}
\frac{(1-\cos \theta )}{2}\Phi =2\pi A(\theta )~R\sin \theta  \label{2}
\end{equation}

and the vector potential

\begin{equation}
\boldsymbol{A}(\theta )=\frac{\Phi }{4\pi R}\frac{(1-\cos \theta )}{\sin
\theta }~\widehat{\phi }  \label{3}
\end{equation}

if we rewrite the vector potential in the Eq. (\ref{3}) into the Eq(\ref{1})
and look for the eigenfunctions in the form $\psi (\theta ,\phi )=T(\theta
)~e^{im\phi }$

\begin{equation}
-\frac{\hslash ^{2}}{2m^{\ast }R^{2}}\left[ \frac{1}{\sin \theta }\frac{d}{%
d\theta }\left( \sin \theta \frac{d}{d\theta }\right) -\frac{m^{2}}{\sin
^{2}\theta }-mp\frac{1-\cos \theta }{\sin ^{2}\theta }-\frac{p^{2}}{4}\frac{%
\left( 1-\cos \theta \right) ^{2}}{\sin ^{2}\theta }\right] T(\theta
)=E~T(\theta )  \label{4}
\end{equation}

where $m\hslash $ is the eigenvalue of $\boldsymbol{L}_{z}$, magnetic flux
quantum $\Phi _{0}=\frac{hc}{e}$ and $\Phi /\Phi _{0}=p$ denotes the
magnetic flux in unit of the flux quantum $\Phi _{0}$. Defining
dimensionless energy $\epsilon =\frac{2m^{\ast }R^{2}}{\hslash ^{2}}E$ and
changing the variable $\mu =\cos \theta $, we write

\begin{equation}
\frac{d}{d\mu }\left[ \left( 1-\mu ^{2}\right) \frac{dT(\mu )}{d\mu }\right]
+\left[ \epsilon -\frac{m^{2}}{1-\mu ^{2}}-mp\frac{1-\mu }{1-\mu ^{2}}-\frac{%
p^{2}}{4}\frac{\left( 1-\mu \right) ^{2}}{1-\mu ^{2}}\right] T(\mu )=0
\label{5}
\end{equation}

Let us define the $T(\mu )$ wave function as:

\begin{equation}
T(\mu )=\left( 1-\mu \right) ^{\frac{\left\vert m\right\vert }{2}}\left(
1+\mu \right) ^{\frac{p+m}{2}}P\left( \mu \right)  \label{6}
\end{equation}

where $p+m\geq 0$ and $m$ can take $-p\leq m\leq p$ integer values. After
substituting Eq. (\ref{6}) into Eq. (\ref{5}), we find

\begin{equation}
\left( 1-\mu ^{2}\right) \frac{d^{2}P(\mu )}{d\mu ^{2}}+\left[
p+m-\left\vert m\right\vert -\left( p+m+\left\vert m\right\vert +2\right)
\mu \right] \frac{dP(\mu )}{d\mu }+\left( \epsilon -\frac{(p+m)(m+\left\vert
m\right\vert +1)+\left\vert m\right\vert }{2}\right) P(\mu )=0  \label{7}
\end{equation}

this differential equation is Jacobi Differential Equation and its solution
are Jacobi Polynomials, $P_{\ell }^{\left( \left\vert m\right\vert
,~p+m\right) }(\mu )$ and $\ell $ is angular momentum quantum number that
nonnegative integer. To find the energy eigenvalues, we use dimensionless
energy identity $\epsilon =\ell \left( \ell +1\right) +\left\vert
m\right\vert (\ell +\frac{1}{2})+(p+m)\left( \ell +\frac{m+\left\vert
m\right\vert +1}{2}\right) $

\begin{equation}
E_{\ell ,p,m}=\frac{\hslash ^{2}}{2m^{\ast }R^{2}}\left[ \ell \left( \ell
+1\right) +\left\vert m\right\vert (\ell +1/2)+(p+m)\left( \ell +\frac{%
m+\left\vert m\right\vert +1}{2}\right) \right]  \label{8}
\end{equation}

This energy eigenvalue is the energy eigenvalue of an electron that freely
moves on the surface of a sphere under the radial magnetic field. And total
wave function

\begin{equation}
\psi (\mu ,\phi )=N_{\ell }^{p,m}\left( 1-\mu \right) ^{\frac{\left\vert
m\right\vert }{2}}\left( 1+\mu \right) ^{\frac{p+m}{2}}~P_{\ell }^{\left(
\left\vert m\right\vert ,~p+m\right) }(\mu )~e^{im\phi }  \label{9}
\end{equation}%
~

from the orthogonality property of Jacobi Polynomials, the normalization
constant $N_{\ell }^{p,m}$ given as:

\begin{equation}
N_{\ell }^{p,m}=\left[ \ell !\frac{\left( 2\ell +p+m+\left\vert m\right\vert
+1\right) ~~\Gamma (\ell +p+m+\left\vert m\right\vert +1)}{\pi
~2^{p+m+\left\vert m\right\vert +2}\Gamma (\ell +\left\vert m\right\vert
+1)\Gamma (\ell +p+m+1)}\right] ^{1/2}  \label{10}
\end{equation}

where $\Gamma (z)$ the Gamma Function. In the absence of a magnetic field,
the wave function is reduced to $\psi (\mu ,\phi )=\frac{2\ell +1}{4\pi }%
~P_{\ell }(\mu )~e^{im\phi }$ where $P_{\ell }(\mu )$ Legendre polynomials.

Now, we examine $p\longrightarrow \infty $ ($p+m\longrightarrow \infty $)
limit. If we change the variable $\mu $ to $x=\frac{p+m}{2}(1-\mu )$,
functions in Eq.(\ref{9}) in this limit $\underset{p+m\longrightarrow \infty
}{lim}P_{\ell }^{\left( \left\vert m\right\vert ,~p+m\right) }(1-\frac{2x}{%
p+m})=L_{\ell }^{\left\vert m\right\vert }(x)$ \cite{Bell1996} where $%
L_{\ell }^{\left\vert m\right\vert }(x)$ Associated Laguerre Polynomials, $%
\left( 1-\mu \right) ^{\frac{\left\vert m\right\vert }{2}}=\left( \frac{2}{%
p+m}\right) ^{\frac{\left\vert m\right\vert }{2}}x^{\frac{\left\vert
m\right\vert }{2}}$ and $\underset{p+m\longrightarrow \infty }{lim}\left(
1+\mu \right) ^{\frac{p+m}{2}}\simeq 2^{\frac{p+m}{2}}~e^{-\frac{x}{2}}$.
Then the wave function becomes

\begin{equation}
\psi (x,\phi )=N_{\ell ,m}~e^{-\frac{x}{2}}~x^{\frac{\left\vert m\right\vert
}{2}}~L_{\ell }^{\left\vert m\right\vert }(x)~e^{im\phi }  \label{11}
\end{equation}

normalization constant $N_{\ell ,m}$ is found from the orthogonality
property of Associated Laguerre Polynomials $N_{\ell ,m}=\left( \frac{\ell !%
}{2\pi ~(\ell +\left\vert m\right\vert )!}\right) ^{\frac{1}{2}}$.

\section{The $R\longrightarrow \infty $ limit}

In this section, we will observe that the wave function and energy belong to
two-dimensional spherical surface under radial magnetic field are converted
to the wave function and energy of an electron under the constant magnetic
field in the z direction on two-dimensional flat surfaces.

In the case of small $\theta $ we can use plane polar coordinates in Eq. (%
\ref{4}), if we take $sin\theta =\rho /R$ and $z\simeq R$, $cos\theta
=1-\rho ^{2}/2R^{2}$ and we take the wave function in the form $\psi (\rho
,\phi )=T(\rho )~e^{im\phi }$

\begin{equation}
\frac{1}{\rho }\frac{d}{d\rho }\left( \rho \frac{dT(\rho )}{d\rho }\right)
+\left( \varepsilon -\frac{m^{2}}{\rho ^{2}}-\frac{p^{2}}{16R^{4}}\rho
^{2}\right) T(\rho )=0  \label{12}
\end{equation}

where $\varepsilon =\frac{2m^{\ast }E}{\hslash
^{2}}-\frac{m~p}{2R^{2}}$. After change of variable
$x=\sqrt{\frac{p}{4R^{2}}}\rho $, we arrive at

\begin{equation}
\frac{1}{x}\frac{d}{dx}\left( x\frac{dT(x)}{dx}\right) +\left( \lambda -%
\frac{m^{2}}{x^{2}}-x^{2}\right) T(x)=0  \label{13}
\end{equation}

where $\lambda =\frac{4R^{2}}{p}\varepsilon $ and considering the limit
states ($x\rightarrow 0$ and $x\rightarrow \infty $) of Eq. (13), we can
propose the solution as

\begin{equation}
T(x)=e^{-x^{2}/2}~x^{\left\vert m\right\vert }~L(x)  \label{14}
\end{equation}

If we substitute Eq. (14) into Eq. (13) and then changing the variable $%
y=x^{2} $, we arrive at

\begin{equation}
y\frac{d^{2}L(y)}{dy^{2}}+\left( \left\vert m\right\vert +1-y\right) \frac{%
dL(y)}{dy}+\left( \frac{\lambda }{4}-\frac{\left\vert m\right\vert +1}{2}%
\right) L(y)=0  \label{15}
\end{equation}

This differential equation is Associated Laguerre differential
equation and its solutions are Associated Laguerre polynomials. To
find energy eigenvalues, $\frac{\lambda }{4}-\frac{\left\vert
m\right\vert +1}{2}=n$ can be used where $n$ is an integer ($n=0,
1, 2,...$). If we use total magnetic flux $\Phi =\int
\boldsymbol{B}.d\boldsymbol{S}=4\pi R^{2}B$
\begin{equation}
E_{n,m}=\hslash \omega _{c}\left( n+\frac{m+\left\vert m\right\vert +1}{2}%
\right)  \label{16}
\end{equation}

where $\omega _{c}=eB/m^{\ast }c$ cyclotron frequency and $B$ constant
magnetic field in $z$ direction on the flat surface. Energy levels in Eq.
(16) is the well-known Landau energy levels \cite{Landau1977}.

\section{Discussions}

In this work, we have exactly solved the Schr\"{o}dinger equation of a
single particle at the spherical surface and under the radial magnetic field
to find electronic energy levels and wave functions. To analyze the energy
levels we have found, consider Eq. (\ref{8}). In the absence of the magnetic field ($p=0$ and $m=0$%
), these energy levels are reduced to $E_{\ell ,0,0}=\frac{\hslash ^{2}}{%
2m^{\ast }R^{2}}\ell \left( \ell +1\right) $ rigid rotator energy levels as
expected. For fixed values of $\ell $ and $p$ in Eq. (\ref{8}), the
difference between energy levels for positive $m$ is $\Delta E=E_{\ell
,p,m+1}-E_{\ell ,p,m}=\frac{\hslash ^{2}}{2m^{\ast }R^{2}}\left( 2\ell
+p+2m+2\right) $. Although the difference between the Landau energy levels $%
\Delta E=\frac{1}{2}\hslash \omega _{c}$ for positive $m$ in flat surfaces
is equidistant, the difference between the energy found for curved surfaces
is $\Delta E=E_{\ell ,p,m+1}-E_{\ell ,p,m}=\frac{\hslash ^{2}}{2m^{\ast
}R^{2}}\left( 2\ell +p+2m+2\right) $ depend on $m$ and is not equidistant.

\begin{figure}[p]
\includegraphics [height=8cm,width=10cm]{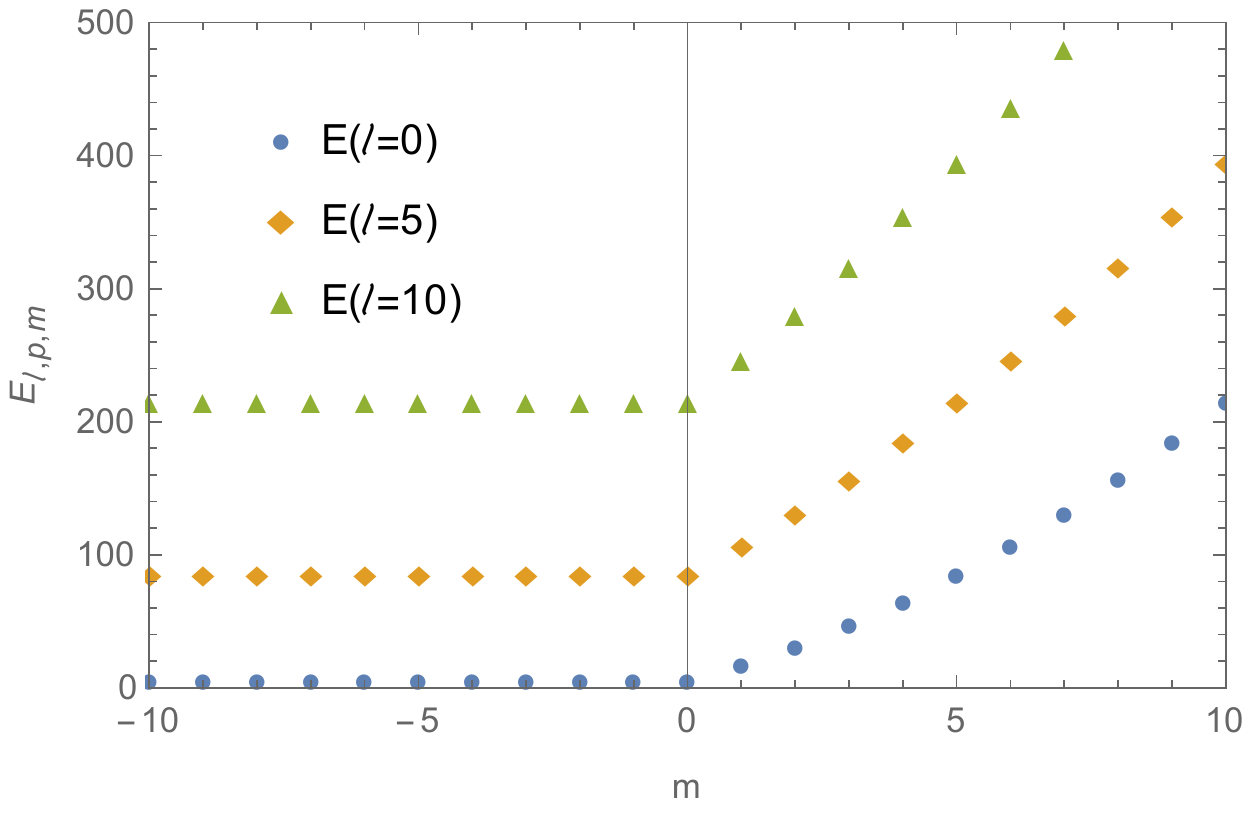}
\caption{Energy levels of an electron on a sphere under the radial magnetic
field as a function of magnetic quantum number $m$ and angular momentum
quantum number $\ell$ for special value of magnetic field (p=10)}
\label{Fig1}
\end{figure}

In Fig. (\ref{Fig1}), we plot energy levels in Eq. (\ref{8}) in
$\left( \frac{\hslash ^{2}}{2m^{\ast }R^{2}}\right) $ unit as a
function of $m$ (magnetic quantum number) and $\ell $ (angular
momentum quantum number) for fixed value of $p=10$. For negative
$m$ values, these energy values are not depend on $m$, but for
positive $m$ values the energy is increased by increasing $m$.
Energy eigenvalues increase with increasing $\ell $ (angular
momentum quantum number) values. In Fig. (\ref{Fig1}), energy levels for $%
\ell =0$, $\ell =5$ and $\ell =10$ denoted by dots, diamonds and triangles,
respectively.

In section 3, it is shown that the energy levels of an electron on the
spherical surface under the radial magnetic field turn into Landau energy
levels at the $R\rightarrow \infty $ limit. Landau energy levels are not
dependent on $m$ (magnetic quantum number) for negative $m$ values but
increase linearly with increasing positive $m$ values as seen from Eq. (\ref%
{16}). The energy levels of an electron on a spherical surface
under the radial magnetic field are similar to those of the Landau
energy levels but there are differences due to spherical surface
effects at the $R\rightarrow \infty $ limit, these spherical
surface effects cease to exist.

\end{document}